\documentclass{ws-procs975x65}

\def\bs{\bigskip}

\begin{document}

\title{NEUTRINO MASS FROM SZ SURVEYS}

\author{YOEL REPHAELI}

\address{School of Physics, Tel Aviv University\\
Tel Aviv 69978, Israel}

\address{CASS, UC San Diego\\
La Jolla, CA 92093\\
yoelr@wise.tau.ac.il}

\author{MEIR SHIMON}

\address{School of Physics, Tel Aviv University\\
Tel Aviv 69978, Israel\\
meirs@wise.tau.ac.il}

\begin{abstract}

The expected sensitivity of cluster SZ number counts to neutrino mass in the sub-eV 
range is assessed. We find that from the ongoing {\it Planck}/SZ measurements the (total) 
neutrino mass can be determined at a ($1\sigma$) precision of $0.06$ eV, if the 
mass is in the range $0.1-0.3$ eV, and the survey detection limit is set at the 
$5\sigma$ significance level. The mass uncertainty is predicted to be lower by a 
factor $\sim 2/3$, if a similar survey is conducted by a cosmic-variance-limited 
experiment, a level comparable to that projected if CMB lensing extraction is 
accomplished with the same experiment. At present, the main uncertainty in modeling 
cluster statistical measures reflects the difficulty in determining the mass 
function at the high-mass end. 
\end{abstract}

\bodymatter

\section{Introduction}

If the sum, $m_\nu$, of all (three) neutrino masses is close to $\sim 0.3$ eV, 
and thus comparable to the energy scale of the recombination epoch, then the 
earliest measurable effect of massive neutrinos on the CMB is their impact on 
the early integrated Sachs Wolfe (ISW) effect. With a mass lower than this value 
neutrinos were a relativistic component that contributed to the decay of linear 
gravitational potentials and thus caused a net change in the temperature of the 
CMB. Measurements of CMB polarization can also constrain neutrino masses 
from measurements of the B-mode lensing-induced signal by the large scale 
structure at redshifts of a few. Ongoing ground-based CMB experiments have 
been searching for this signal at its expected peak ~$l\sim 1000$, on smaller 
angular scales than the predicted, much weaker primordial (inflation-induced) 
B-mode signal which is expected to peak around $l\sim 100$.

It has been conjectured that employing optimal estimators to CMB temperature 
and polarization maps obtained from full-sky measurements with a 
cosmic-variance-limited (CVL) experiment will allow recovering the lensing 
potential with precision that could constrain $m_\nu$ at a level of $\sim 0.04$ eV 
\cite{kap}.
This projection was based on the assumption 
that foregrounds are negligible and there is no source of non-gaussianity other 
than CMB lensing, and therefore it is likely to be overly optimistic. Analysis 
of the WMAP7 database (with BAO and $H_0$ priors) yielded the upper 
limit \cite{kom} 
$m_\nu < 0.58$ eV (95\% CL). 
Other cosmological probes of neutrino masses include weak lensing shear 
maps \cite{coo,aba,nam},
galaxy \cite{hu,tho}, and 
Ly$\alpha$ surveys \cite{cro,gra}.

To assess the relative importance of the yields of these various cosmological 
probes, their respective upper limits have to be compared with lower limits 
from neutrino oscillation experiments. Measured mass values from neutrino 
oscillations imply that at least one of the neutrinos is 0.05 eV or heavier, 
if the mass hierarchy is 'normal', whereas in the 'inverted' hierarchy two 
neutrino masses are each above 0.05 eV. Thus, the lowest bound on $m_\nu$ is 
in the range 0.05-0.10 eV, which sets the benchmark for determining the 
hierarchy and possibly ruling out one of these hierarchies. This sets the 
goal for cosmological neutrino mass precision to better than 0.05 eV.

The SZ effect is a unique probe of cluster and cosmological parameters; its 
statistical diagnostic value is gauged by cluster number counts and the power 
spectrum of the CMB anisotropy it induces. The steep dependence of SZ number 
counts on $\sigma(M,z)$ - the rms mass fluctuation on mass scale $M$ at 
redshift $z$ - which depends exponentially on $m_\nu$ (e.g., 
Ref. ~\refcite{les1}), 
makes cluster surveys sensitive probes of neutrino mass, as has already been 
demonstrated \cite{wan,ssr,car}. 

In this paper we summarize results from our recent analysis \cite{sri} 
in which we extended and improved our earlier predictions \cite{ssr}
of the precision with which $m_\nu$ can be determined from SZ measurements 
by the ongoing {\it Planck} and future CVL surveys. Our approach and more details 
of the Fisher matrix analysis are only very briefly discussed here; a more 
detailed description is given in Refs.~\refcite{ssr,sri}. 

\section{LSS Evolution with Massive Neutrinos}

Evolution of the large scale structure in the matter-dominated era is described 
in terms of the matter power spectrum,
\begin{eqnarray}
P_{m}(k,z)=Ak^{n}T^{2}(k,z),
\end{eqnarray}
where $Ak^{n}$ is the primordial density fluctuation spectrum with index $n$, 
and normalization $A$;  $T(k,z)=T(m_{\nu};k,z)$ is the transfer function. 
Normalization of the power spectrum measured at present is in terms of the mass 
variance parameter on a scale of $R={\rm 8 \,Mpc\ h^{-1}}$, 
\begin{eqnarray}
\sigma_{8}^{2}=\int_{0}^{\infty}P_{m}(k,z)W^2(kR)k^2\frac{dk}{2\pi^{2}},
\end{eqnarray}
where $W(kR)$ is a window function (typically assumed to be top-hat). 
The essence of neutrino impact on density fluctuations can be appreciated from the 
fact that for masses $\leq 1$ eV, diffusion damping of density fluctuations is on 
scales below a few tens of Mpc. The suppression of these scales is respresented in 
the transfer function. It is expected that $m_{\nu}$ and $\sigma_{8}$ are 
anti-correlated, and due to the steep dependence of cluster counts on 
$\sigma_{8}$, we expect also strong dependence on $m_{\nu}$.

The total neutrino mass can be deduced from a comparison of the observed number 
of clusters in a given redshift bin to the number predicted from the mass 
function, $\frac{dn(M;z)}{dM}$, which is defined in terms of the differential 
number of clusters in a volume element $dV$,
\begin{eqnarray}
dN(M,z)=f_{sky}\frac{dn(M,z)}{dM}dVdM ,
\end{eqnarray}
where $f_{sky}$ is the observed sky fraction ($0.65$ for the two experiments 
considered here). The number of clusters in a given interval $\Delta z$ around $z_{i}$ is
\begin{eqnarray}
\Delta N(z_{i})=f_{sky}\Delta z_{i}\frac{dV(z_{i})}{dz}\int\frac{dn(M,z_{i})}{dM}dM.
\end{eqnarray}

In the analysis described here we used the code in Ref.~\refcite{kia} 
to calculate $T(m_{\nu};k,z)$, but we did use the default transfer function 
in (the CMB analysis program) CAMB to calculate the primordial angular power 
spectrum of the CMB. This procedure is justified given that the main 
difference between the two transfer functions is only apparent at large 
values of $k$.

The Tinker et al. (2008) mass function \cite{tin}, which was obtained from a 
large set of dynamical cosmological simulations in the $\Lambda$CDM model, 
was adopted. With the mass function expressed in the usual form, 
\begin{eqnarray}
\frac{dn}{dM}=f(\sigma)\frac{\rho_{m}}{M}\frac{d\ln(\sigma^{-1})}{dM} ,
\end{eqnarray}
the analytic approximation \cite{tin} is 
\begin{eqnarray}
f(\sigma)=A\left[\left(1+\frac{\sigma}{b}\right)^{-a}\right]
e^{-\frac{c}{\sigma^{2}}} .
\end{eqnarray}
The parameters $A$, $a$, $b$ and $c$, which depend on $z$ and the overdensity 
at virialization, $\Delta_{v}$, are
\begin{eqnarray}
A&=&A_{0}(1+z)^{-0.14}\nonumber\\
a&=&a_{0}(1+z)^{-0.06}\nonumber\\
b&=&b_{0}(1+z)^{-\alpha}\nonumber\\
\log(\alpha)&=&-\left(\frac{0.75}{\log(\Delta_{v}/75)}\right)^{1.2}\nonumber\\
b_{0}&=&1.0+(\log(\Delta_{v})-1.6)^{-1.5}
\end{eqnarray}
where $c$ and $A_{0}$ were obtained from Table 2 of Tinker et al. (2008). 
Values listed in the table were used also for deriving fits for $a_{0}$ and 
$b_{0}$ as functions of $\Delta_{v}$,  
\begin{eqnarray}
a_{0}&=&1.7678\alpha_{1} -0.5941\alpha_{2}\exp(-0.02924\Delta_{v}^{0.5967})\nonumber\\
c&=&1.7077\alpha_{3} -0.7038\alpha_{4}\exp(-0.001\Delta_{v}^{1.079})
\end{eqnarray}
where we introduced the additional four (`nuisance') parameters 
$\alpha_{1}-\alpha_{4}$ in order to account for possible uncertainties 
(UCs) or biases in the Tinker et al. parameters.

Cluster DM profiles were approximated by the NFW model, with a 
mass-concentration relation $c(M,z)$ taken from Ref.~\refcite{duf}.
Intracluster (IC) gas was assumed to be well described by a polytropic 
equation of state with index $\Gamma=1.2$. The solution of the equation of 
hydrostatic equilibrium for a polytropic gas inside the potential well of a 
DM halo is \cite{ost} 
\begin{equation}
 \rho(x)=\rho_0\left[1-\frac{B(\Gamma-1)}{\Gamma}\left(1-
\frac{\ln(1+x)}{x} \right)\right]^{1/(\Gamma-1)}
\label{eq:rhox}
\end{equation}
where $x=r/r_s$, $r_s$ is the scale factor of the NFW density profile, $B$ is
given by $B=4\pi G\rho_s r_s^2\mu m_p/k_B T_0$; $\mu$ is mean molecular 
weight, and $m_p$ is the proton mass. More details on the IC gas model are 
given in Ref.~\refcite{drs}.

The mass fraction of IC gas was assumed to have the scaling deduced 
in Ref.~\refcite{vik},
\begin{eqnarray}
f_{g}(M,z)=\alpha_{5}[0.125+0.037\log_{10}(M_{500}/{10^{15}M_\odot})]
\end{eqnarray}
with an added parameter $\alpha_{5}$ (whose fiducial value is $1$) to account 
for UCs or biases in this scaling relation. $M_{500}$ is the total 
cluster mass within a sphere whose mean density is 500 that of the background. 
The $z$-dependence in the last equation is that of $M_{*}$, with $M_{*}$ 
defined such that for a fixed redshift the mass fluctuation is 
$\sigma(M_{*},z)=1.686$.
  
The SZ power spectrum was normalized to the value measured \cite{rei} by SPT, 
$C_{l}=3.65 \,\mu K^{2}$ at $l=3000$. 
This was done separately for each fiducial value of $m_\nu$. We note that 
even though the shape of the SZ power spectrum depends on the fiducial 
neutrino mass, power levels are nearly the same for $\ell \sim 2000 - 3000$. 
Consequently, a similar number of clusters is expected to be detected, 
implying that neutrino mass UC are essentially independent on the assumed 
fiducial neutrino mass.

\section{Likelihood Analysis}

We constructed a likelihood function for cluster number counts and carried out 
the diagnostic analysis based on calculations of Fisher matrices for the primary 
CMB with and without lensing extraction, employing the standard approach (e.g., 
Refs.~\refcite{les2,ssr})
which we do not describe here.

In the context of the flat $\Lambda$CDM model the global parameters were the 
normalization $A$, the power-law index $n$ of the primordial scalar perturbations, 
density parameters of matter, $\Omega_{m}$, and baryons, $\Omega_{b}$, dark energy 
equation of state parameter, $w$, the Hubble parameter (scaled to $100$ 
km/sec/Mpc) $h$, primordial helium abundance $Y_{p}$, optical depth to 
reionization $\tau$, and the neutrino mass $m_{\nu}$. Additionally, we adopted 
priors on the cosmological parameters obtained from the primary and lensed sky 
observed with {\it Planck} and a CVL experiment, with the corresponding Fisher matrices 
denoted by $F^{P}$ and $F^{LE}$, respectively. We also used the prior 
$H_{0}=71.0\pm 2.5$ km/sec/Mpc. In addition to the nine global parameters we 
introduced four parameters to account for possible UCs in parameter 
values of the Tinker et al. (2008) mass function, Eq. (8), and a parameter to 
account for UC in the gas mass fraction, Eq. (10).
 
Our likelihood function for cluster number counts is based on a Poisson 
distribution for the observed and expected number counts in redshift-shells 
(e.g., Ref.~\refcite{hol}).
While for high cluster abundances a spatial cluster correlation 
term would generally be included, the selection of a high $5\sigma$ detection 
threshold ensures that only the most massive clusters are relevant for our 
analysis, effectively minimizing the impact of the relatively small correlation 
term, whose contribution to our results for the neutrino mass UC we estimate 
to be an increase of up to $\sim 15\%$. Cluster counts were calculated 
in redshift shells with width $\Delta z=0.1$ up to $z=1$. This width is larger 
than predicted photo-z redshift UCs which are at the 
$\sigma_{z}=0.02(1+z)$ level; we verified that further refinement of the 
redshift bins did not affect the results. The choice of maximal redshift was 
based on the realization that the strength of the neutrino mass constraint is 
largely determined by low-redshift, high-mass clusters. 

We calculated number counts in redshift bins rather than in  redshift and 
mass bins since mass `slicing' was found not to enhance the diagnostic power 
due to the requirements that each cell contains at least 20 clusters, the need 
to select wide mass bins to account for the large UC (of a factor of $\sim 2$) 
in mass determination, and the fact that mass and redshift of clusters that 
satisfy the high detection threshold are strongly correlated. Thus, cluster 
counts in redshift bins is obviously preferable to using the mass as an 
explicit parameter due to the fact that redshift is well defined and precisely 
measured.

Predicted precision in constraining $m_\nu$ is calculated from the Poissonian 
likelihood function for number counts in the $i$'th redshift bin 
\begin{eqnarray}                                                                
\mathcal{L}_{i}(N_{i}^{p};N_{i}^{o})                                            
\propto\frac{(N_{i}^{p})^{N_{i}^{o}}\exp(-N_{i}^{p})}{N_{i}^{o}!} ,             
\end{eqnarray}                                                                  
where $N_{i}^{p}$ is the predicted (given a set of cosmological parameters and 
$m_\nu$) and $N_{i}^{o}$ the observed cluster number in the $i$'th redshift bin. 
Since $N_{i}^{p}$ is a function of several parameters $\lambda_{k}$, small 
deviations with respect to its expected fiducial value can be determined from 
$N_{i}^{o}\approx N_{i}^{p}+\sum_{k}\frac{\partial N_{i}^{p}}
{\partial\lambda_{k}} \Delta\lambda_{k}$. The Fisher matrix for number counts 
is then calculated from the likelihood function 
\begin{eqnarray}                                                                
F_{jl}^{N}=-\frac{\partial^{2}\mathcal{\ln L}}{\partial\lambda_{j}\partial
\lambda_{l}} 
=\sum_{i}\frac{1}{N_{i}}\frac{\partial N_{i}}                                   
{\partial\lambda_{j}}\frac{\partial N_{i}}{\partial\lambda_{l}}.                
\end{eqnarray}                                                   
The estimated UC in the parameter $\lambda_{j}$ is the square root of the 
respective Fisher matrix element,                  
\begin{eqnarray}                                                                
\Delta\lambda_{j}=(F_{jj}^{N})^{-1/2} .
\end{eqnarray}                                                                  

The full Fisher matrix is a sum of the number counts, $F_{jl}^{N}$, and 
either the primary CMB, $F_{jl}^{P}$, or the lensed CMB, 
$F_{jl}^{{\rm LE}}$. For the calculation of the signal-to-noise 
$\mathcal{S/N}$ with which a cluster can be detected in a survey we assumed 
that main sources of noise are instrumental, primary CMB anisotropy, and point 
source contamination. The performance of optimal matched filters (as applied 
in Ref.~\refcite{ssr}) 
was assumed in order to estimate the abundance of detected clusters on a 
finely-sampled $M-z$ grid.

\section{Projected $m_\nu$ Precision Levels}

The standard $\Lambda$CDM cosmological model was assumed  with 
WMAP-7 best-fit parameters, but no priors were set on the cosmological 
parameters, except for $H_{0}$ for which an UC of 2.5 km/sec/Mpc 
was assumed. As we specified in the previous section, the cluster population 
was described in terms of the Tinker et al. (2008) mass function with four 
additional parameters ($\alpha_{1}-\alpha_{4}$) that gauge the robustness 
of  $\sigma_{m_{\nu}}$ to UCs in the mass function. The calculated 
counts included all clusters in the mass range 
$3\times 10^{13}M_{\odot}-3\times 10^{15}M_{\odot}$. We verified that the 
high $\mathcal{S/N}$ detection threshold we set guaranteed that the 
detected clusters are actually much more massive than the low mass 
end of this range. We used the scaling relation deduced in Ref.~\refcite{vik} 
for the gas mass fraction in clusters, but parametrized it with an 
added multiplicative factor. 

\begin{table}
Table 1: {\it Planck} sky coverage and sensitivity parameters. 
Channel sensitivities and beam (FWHM) sizes are taken from Table 4 of 
Ref. ~\refcite{les1}, where sensitivities of polarization measurements 
(in the first seven channels) are also listed. Only the 100, 143 and 353 
GHz channels were used in the computation of number counts; see the 
text for details. 
\bs

\centering
\begin{tabular}{|c|c|c|c|c|}
\hline
$f_{\rm sky}$& $\nu [GHz]$ & $\theta_b [1']$ & $\Delta_T [\mu K]$\\
\hline
\hline
&  30 & 33 &  4.4\\
&  44 & 23 &  6.5\\
&  70 & 14 &  9.8\\
&  100 & 9.5 &  6.8\\
0.65& 143 & 7.1 & 6.0\\
&  217 & 5.0 &  13.1\\
& 353 & 5.0 & 40.1\\
&  545 & 5.0 &  401\\
& 856 & 5.0 & 18300\\
\hline
\end{tabular}
\end{table}

For simulating cluster detection by {\it Planck} all nine frequency channels were 
used in the calculation of $F^{P}$ and $F^{LE}$ , whereas only the 100, 143 
and 353 GHz channels were used in our calculations of the number counts. 
The relevant {\it Planck} specifications are listed in Table 1. Results for the 
projected neutrino mass UC from analyses of cluster number 
counts from the (ongoing) {\it Planck} survey and a similar (future) CVL 
survey are presented in Table 2.  Listed in the table (from left to right) are 
the calculated UC in $m_{\nu}$ from the primary (P) CMB (both 
temperature and polarization anisotropy), lensing extraction (LE) of the CMB, 
primary CMB and number counts, and finally lensed CMB and number counts. 

\begin{table}
Table 2: Statistical UC on total neutrino mass from cluster number counts 
obtained from the {\it Planck} and CVL SZ surveys; see the text for details. 
\bs

\centering
\begin{tabular}{|c|c|c|c|c|c|c|}
\hline
Survey&$\alpha_1$, $\alpha_2$, $\alpha_3$, $\alpha_4$&$\sigma_{m_{\nu}}[eV]$&$\sigma_{m_{\nu}}[eV]$&$\sigma_{m_{\nu}}[eV]$&$\sigma_{m_{\nu}}[eV]$&$N$\\
&UC [\% ]&P&LE&P+N(z)]&LE+N(z)&\\
\hline\hline
&0&&&0.06&0.06&\\
{\it Planck}&3&0.43&0.15&0.07&0.06&6040\\
&5&&&0.08&0.07&\\
&10&&&0.12&0.09&\\
\hline\hline
&0&&&0.04&0.03&\\
CVL&3&0.29&0.05&0.06&0.04&13860\\
&5&&&0.07&0.04&\\
&10&&&0.11&0.05&\\
\hline
\end{tabular}
\end{table}

The values of the neutrino mass uncertainty listed in Table 2 clearly 
demonstrate that cluster number counts alone (but with priors from measurements 
of the primary CMB power spectrum and the HST prior on $H_{0}$) neutrino 
mass uncertainties may be constrained to the $\sim 0.04-0.06$ eV range, 
depending on the value of  $m_\nu$ and the nature of the survey. 

The projected neutrino mass precision for {\it Planck} and a CVL experiment are based 
on the Tinker et al. (2008) analytic fit to their simulations. To test the 
robustness of our estimates to possible deviations from values of the parameters 
in their analytic representation, we allowed for $1-10\%$ uncertainty in each 
of the (added) parameters $\alpha_{1}-\alpha_{4}$ (around their fiducial value 
of $1$). Doing so increases the neutrino mass uncertainty, $\sigma_{m_{\nu}}$ 
from $0.06$ eV to $0.12$ when the database consists of  the primary CMB and 
{\it Planck} number counts. Repeating the analysis with the primary CMB and number 
counts expected from a CVL survey, the corresponding $\sigma_{m_{\nu}}$ 
changes from $0.04$ to $0.11$ eV. Accounting for the inherent uncertainty in 
the mass function (especially at the high mass end) is clearly very important 
for placing reliable constraints on the neutrino mass. 

\section{Summary}

The primary CMB anisotropy (including the lensing signature imprinted 
by the large scale structure) is not an optimal probe of processes and 
phenomena on ${\rm \sim Mpc}$ scales. Structure on ${\rm \sim Mpc}$ 
scales probes the entire evolutionary history of matter perturbations 
down to these scales. This is especially relevant to neutrino physics 
via the effect of neutrino free streaming on these and larger scales. 
Free streaming of neutrinos with masses O($0.1$) eV affects the matter 
power spectrum on the characteristic scales of galaxy clusters; this is 
what makes cluster number counts a more optimal probe of $m_\nu$ in 
this mass range. 

For the observationally-allowed range $m_{\nu} \sim 0.1-0.3$ eV, the projected 
uncertainty in $m_\nu$ is relatively small, in the range $\sim 0.04-0.06$ eV, a 
range that is competitive with predicted results form CMB lensing extraction. 
We conclude that our results provide strong motivation for performing the analysis 
described here with actual (rather than projected) ongoing {\it Planck} survey data, 
and - if still relevant - with future results from a CVL experiment.

As we have noted, the most important source of uncertainty in modeling the 
evolution of clusters in mass and redshift is the mass function. Estimated 
uncertainties in this basic function were explicitly included in our analysis. 
Extensive cosmological hydrodynamical simulations are expected to provide 
a more accurate description of the population, and significantly improved 
sampling of cluster abundance particularly at the high-mass end. Uncertainties 
in modeling IC gas and the evolution of the gas mass fraction are also relevant.

\section*{Aknowledgements}

This work was supported by the James B. Ax Family Foundation, the 
TAU-UCSD Cosmology Program, and by grants from the US-Israel BSF (2008452) 
and the Israel Science Foundation (1496/12).

\end{document}